\documentclass[aps,prl,twocolumn,superscriptaddress,amsmath,amssymb,floatfix]{revtex4-2}
\usepackage{graphicx}
\usepackage{bm}

\providecommand{\aap}{Astron. Astrophys.}

\providecommand{\apj}{Astrophys. J.}
\providecommand{\apjl}{Astrophys. J. Lett.}

\providecommand{\aj}{Astron. J.}
\providecommand{\mnras}{Mon. Not. R. Astron. Soc.}
\providecommand{\nat}{Nature}

\providecommand{\prd}{Phys. Rev. D}
\providecommand{\prl}{Phys. Rev. Lett.}

\begin{document}

\title{The angular structure of the GW170817 jet from prompt emission alone}

\author{R.~Moradi}
\email{rahim.moradi@icranet.org}
\affiliation{ICRANet, P.zza della Repubblica 10, I--65122 Pescara, Italy}
\affiliation{ICRA and Dipartimento di Fisica, Sapienza Universit\`a di Roma,
P.le Aldo Moro 5, I--00185 Rome, Italy}
\author{R.~Ruffini}
\affiliation{ICRANet, P.zza della Repubblica 10, I--65122 Pescara, Italy}
\affiliation{ICRA and Dipartimento di Fisica, Sapienza Universit\`a di Roma,
P.le Aldo Moro 5, I--00185 Rome, Italy}
\affiliation{INAF, Viale del Parco Mellini 84, I--00136 Rome, Italy}

\date{\today}

\begin{abstract}
We determine the angular structure of the GW170817 jet by the prompt emission alone, without afterglow fitting, circumburst density or microphysical parameters. We assume that GRB 090510 and GW170817 have outflows of the same kind, observed respectively on-axis and at the interferometric viewing angle of $20^\circ$. We support this assumption with independent gravitational-wave data showing compatible binary masses and radiated energies. We show that for an observer whose beaming cone is filled with outflow, the point-source Doppler scalings do not apply: $E_{\rm iso}=4\pi\epsilon(\theta_v)$, with $\epsilon$ the energy radiated per unit solid angle along the line of sight, while the peak energy follows $E_{\rm p,i}\propto\Gamma(\theta_v)$. We obtain $n={\rm d}\ln\epsilon/{\rm d}\ln\Gamma = 3.76\pm0.29$ from the ratio of the two bursts, with no free parameter and no assumed angle. This excludes four structures in common use at $4.7\sigma$ to $13\sigma$; three remain above $4\sigma$ across the full reported range of the peak energy of the GW170817 jet. Two prompt spectra fix no angular scale; supplying it with the core Lorentz factor of GRB 090510 and the viewing angle gives $\epsilon\propto\theta^{-7.4}$ outside a core of $2^\circ$--$5^\circ$, in agreement with the width inferred from $367$ short bursts, the outflow remaining relativistic at $\Gamma=33$ on the line of sight. The exponent exceeds what the Lorentz boost of a uniform comoving flow can produce, so the structure is intrinsic to the outflow and not a consequence of the boost. The same structure fixes the emission radius, which contributes $0.41$~s of the $1.74$~s delay between the gravitational-wave signal and the gamma-rays, the remainder being the launch and breakout of the jet, with no free parameters. We conclude that the faintness lies in the structure of the GW170817 jet, not in the de-beaming of a bright core.
\end{abstract}

\maketitle

The prompt electromagnetic counterpart of GW170817, here named the GW170817 jet, is approximately six orders of magnitude fainter than the short gamma-ray bursts with which it is classified \cite{2017ApJ...848L..13A}. Interferometry has established that its outflow was relativistic and seen off axis \cite{2018Natur.561..355M,2019Sci...363..968G, 2022Natur.610..273M}. However, geometry does not provide the angular structure of the outflow, the dependence on polar angle of the energy radiated per unit solid angle, $\epsilon(\theta)$, and the bulk Lorentz factor, $\Gamma(\theta)$. In fact, structure is what converts an observed brightness into a true energy release, which is clustered near $5\times10^{50}$~erg once the collimation is taken into account \cite{2001ApJ...562L..55F}. Moreover, a) it sets the fraction of bursts whose emission reaches us, b) it allows a single intrinsic configuration to account for the spread in observed luminosity \cite{2002MNRAS.332..945R,2002ApJ...571..876Z}, and c) it carries the imprint of how the outflow was collimated by the engine that launched it \cite{1977MNRAS.179..433B}. These are obtained by fitting the panchromatic afterglow over months \cite{2019Sci...363..968G,2020A&A...641A..61A}, which requires prior knowledge of circumburst density, microphysical parameters, and shock dynamics, none of which is known independently.

Here we determine the outflow structure from the prompt gamma-ray emission alone. The rigorous
results obtained in the analysis of GRB 220101A, which have found coincidences with historical
astrophysical events, have convinced us of a previously unexpected conclusion: that rigorously
derivable results have general validity, and can be interchanged simply by using the
appropriate boundary conditions \cite{2026arXiv260820829R}. This is assumed here for the
binary neutron star merger of GRB 090510 and the gravitational-wave emission of GW170817. We
treat GRB 090510 and GW170817 as the same kind of astrophysical system observed from two
different lines of sight. This approach requires at least two such bursts, one of them observed sufficiently well that its Lorentz factor is measured empirically rather than assumed. GRB 090510 provides this latter condition: a $30.5$~GeV photon constrains its bulk Lorentz factor to $\Gamma\gtrsim1200$ through $\gamma\gamma$ transparency \cite{2010ApJ...716.1178A}. Consequently, we adopt GRB~090510 as the on-axis member of the pair, with GW170817 as the off-axis counterpart.

What follows is established to three different degrees, which we keep separate. The first is a point of principle: the point-source Doppler scalings fail for an observer whose beaming cone is filled with outflow, regardless of the properties of these two bursts. The second is our measurement, the slope of the radiated energy against the Lorentz factor. It needs the principle, the premise that the two systems are of the same kind, and four observed values, and it needs nothing else. The third is the structure written as a run with polar angle. It needs all of these, and two quantities in addition, which we take from elsewhere: the core Lorentz factor of GRB 090510 \cite{2010ApJ...716.1178A} and the viewing angle ($\theta_v$) of GW170817 measured by Mooley \textit{et al.} \cite{2018Natur.561..355M,2022Natur.610..273M}; see Table~\ref{tab:in}. In Table~\ref{tab:in}, we also report the slope of the kinetic energy per unit solid angle from the afterglow treatment of Ghirlanda \textit{et al.} \cite{2019Sci...363..968G}, which is needed for calculating the radiative efficiency.

We take the core Lorentz factor of GRB 090510 from the compactness limit, which assumes that the gamma-rays are emitted in a single zone. However, \citet{2018PTEP.2018d3E02I} have questioned this assumption, showing that in multi-zone models the limit is reduced by a factor of several, from $\Gamma_c\gtrsim1200$ to $\Gamma_c\approx300$. In the Supplemental Material \cite{SM} we show that our result does not depend on which of the two is right.

\begin{table}[t]
\caption{\label{tab:in}What the measurement uses. The exponent $n$ requires only the four
observables in the upper block. The anchors convert it into a run with polar angle; the
afterglow slope enters the efficiency alone.}
\begin{ruledtabular}
\begin{tabular}{lll}
quantity & value & source \\
\hline
$E_{\rm iso}$ (090510) & $(3.95\pm0.21)\times10^{52}$~erg & \cite{2016ApJ...832..136R} \\
$E_{\rm iso}$ (GW170817) & $(5.3\pm1.0)\times10^{46}$~erg & \cite{2017ApJ...848L..13A} \\
$E_{\rm p,i}$ (090510) & $7.89\pm0.76$~MeV & \cite{2016ApJ...832..136R} \\
$E_{\rm p,i}$ (GW170817) & $0.217\pm0.055$~MeV & \cite{2017ApJ...848L..14G} \\
\hline
$\Gamma_c$ (090510) & $\gtrsim1200$ & \cite{2010ApJ...716.1178A} \\
$\theta_v$ (GW170817) & $19^\circ$--$25^\circ$ & \cite{2018Natur.561..355M,2022Natur.610..273M} \\
${\rm d}E_{\rm kin}/{\rm d}\Omega$ & $\propto\theta^{-5.5}$ & \cite{2019Sci...363..968G} \\
\hline
main pulse, rest frame & $0.158$, $0.570$~s & \cite{2017ApJ...848L..14G} \\
GW--GRB delay & $1.74\pm0.05$~s & \cite{2017ApJ...848L..13A} \\
\end{tabular}
\end{ruledtabular}
\end{table}

The assumption of a common outflow is not established by the gamma-ray data alone, which are what the premise is used to interpret. We therefore turn to the gravitational waves, a channel the rest of this work does not use. The total energy a binary radiates in gravitational waves is a property of the system and does not depend on the direction from which it is seen; the energy per unit solid angle does vary with inclination, but by no more than a factor of $8$ between the polar and the equatorial direction \cite{1963PhRv..131..435P}. Two binaries of the same kind can therefore be compared without considering the angle from which each is observed. The same treatment of GRB 090510 bounds the inspiral release of its $2.38\,M_\odot$ binary at $7.6\times10^{52}$~erg \cite{2026JHEAp..5000464R}. For GW170817, of total mass $2.74\,M_\odot$, the LIGO--Virgo analysis places a lower limit on the energy emitted below $600$~Hz, $E_{\rm rad}>0.025\,M_\odot c^2=4.5\times10^{52}$~erg \cite{2017PhRvL.119p1101A}. The reported energy carries no angular information. The two limits are compatible and the binaries are of similar mass and nature, while the prompt gamma-rays of the two bursts differ by approximately six decades. The strain itself cannot provide the viewing angle, so the angular anchor is taken from the interferometry \cite{SM}.

\begin{figure}[t]
\includegraphics[width=\columnwidth]{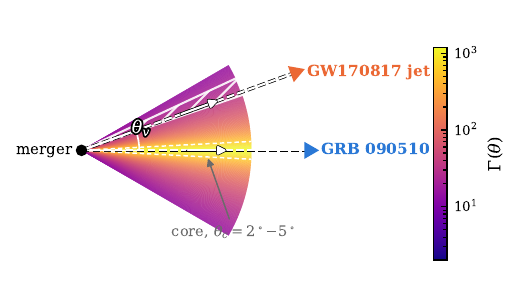}
\caption{\label{fig:geom}The geometry, to scale in polar angle. Colour gives the bulk Lorentz factor for a core half-angle $\theta_c=3^\circ$ and profile index $a=1.90$. Both observers lie \textit{inside} the outflow, each receiving light from material on its own line of sight that moves head-on (white arrows), and for GW170817 the core is beamed away entirely. The hatched band is the interferometric viewing angle, $19^\circ$--$25^\circ$.} 
\end{figure}

\textit{The transformation--Boost.}---The polar angle from the symmetry axis of the outflow is $\theta$, and $\theta_v$ is the angle between that axis and the line of sight. We define $\epsilon(\theta)$ and $\Gamma(\theta)$ as the energy radiated per unit solid angle, and the bulk Lorentz factor, respectively.  For a geometrically thin, optically thin shell of isotropic comoving emissivity, expanding radially, whose radius does not change appreciably during the emission, the isotropic-equivalent energy for an observer at $\theta_v$ is \cite{2015MNRAS.450.3549S}
\begin{equation}\label{eq:master}
E_{\rm iso}(\theta_v)=\int\frac{\delta^3(\theta,\phi;\theta_v)}{\Gamma(\theta)}\,
\epsilon(\theta)\,{\rm d}\Omega.
\end{equation}

The integration runs over the outflow, ${\rm d}\Omega=\sin\theta\,{\rm d}\theta\,{\rm d}\phi$ and $\phi$ is the azimuth of the radiating element about the axis, measured from the half-plane which contains the line of sight. The radiating element possesses the Doppler factor $\delta=\{\Gamma(\theta)[1-\beta(\theta)\cos\psi]\}^{-1}$, where $\beta=(1-\Gamma^{-2})^{1/2}$ is its velocity in units of $c$ and $\psi$ is the angle that velocity makes with the line of sight. The expansion being radial, $\psi$ is also the angle between the direction of the element and the line of sight, so that $\cos\psi=\cos\theta\cos\theta_v+\sin\theta\sin\theta_v\cos\phi$.  Matter lying on the line of sight has $\psi=0$ and $\delta=\Gamma(1+\beta)\simeq2\Gamma(\theta_v)$.

Assigning the whole burst the Doppler factor on the line of sight leads to the point-source scaling $E_{\rm iso}\propto\delta^3$, or $\delta^4$ at fixed radiating time \cite{2018pgrb.book.....Z}. This holds only when the emitting matter subtends less than the beaming cone, and it does not hold here: both observers lie inside the outflow (Fig.~\ref{fig:geom}). \citet{2019Sci...363..968G} draw the same inference for this burst, that its gamma-rays come from the
sheath moving toward the observer and not from the core, whose emission would be beamed too narrowly to reach it. Treatments of the prompt emission of
GRB 090510 and of other bursts approximate the expanding plasma as spherically symmetric with a single Lorentz factor at transparency
\cite{2026JHEAp..5000464R,2021PhRvD.104f3043M,2018ApJ...869..151R}; that geometry already places the observer within the outflow, and what changes here is not how far the outflow extends, but the assumption that $\Gamma$ is the same in every direction. 

If $\Gamma$ were angle independent, the kernel would integrate exactly over the whole sphere, $\int(\delta^3/\Gamma)\,{\rm d}\Omega=4\pi$, leading
\begin{equation}\label{eq:Eiso}
E_{\rm iso}(\theta_v)\longrightarrow 4\pi\,\epsilon(\theta_v) .
\end{equation}

The boost of each element is cancelled by the $\Gamma^{-2}$ shrinking of the region an observer can see. The outflow treated here is not uniform. What the limit requires of it is not uniformity but smoothness: that $\epsilon$ and $\Gamma$ vary little across the patch of width $1/\Gamma(\theta_v)$ about the line of sight. The departure is second order in the ratio of that width to the angular scale on which the profile varies. An off-axis observer of a structured jet then sees the local emission and not a de-beamed core, and the weakness of the GW170817 jet is a property of $\epsilon$ at $\theta_v$ rather than of the boost. We do not rely on the limit in any case: Eq.~(\ref{eq:master}) is integrated as it stands throughout, and the departures are $0.04\%$ on axis and $1.9\%$ at $\theta_v$ (Supplemental Material \cite{SM}, Sec.~I).

The spectral peak is set by the typical blueshift rather than by an energy budget. Evaluating the structured-jet spectrum of \cite{2015MNRAS.450.3549S} for the profile obtained below, the peak of $\nu\mathcal{F}_\nu$ falls at $1.4753\,\Gamma\nu_0'$ on axis and $1.4747\,\Gamma\nu_0'$ at $\theta_v=20^\circ$, so
\begin{equation}\label{eq:Ep}
E_{\rm p,i}(\theta_v)\propto\Gamma(\theta_v)
\end{equation}
to better than $0.1\%$ between the two bursts.

\textit{The exponent.}---Equations~(\ref{eq:Eiso}) and (\ref{eq:Ep}) make the isotropic energy of a burst the energy radiated per unit solid angle along its own line of sight and its peak energy proportional to the Lorentz factor there. For two bursts $A$ and $B$ the left-hand sides are measured and the right-hand sides are properties of the outflow at the two lines of sight. Defining
\begin{equation}\label{eq:def}
\frac{E_{\rm iso}^{(A)}}{E_{\rm iso}^{(B)}}
=\left(\frac{E_{\rm p,i}^{(A)}}{E_{\rm p,i}^{(B)}}\right)^{n}
\quad\Rightarrow\quad
n=\left\langle\frac{{\rm d}\ln\epsilon}{{\rm d}\ln\Gamma}\right\rangle ,
\end{equation}
so that $n$ is the mean logarithmic slope of the radiated energy against the Lorentz factor over the range the two bursts span. It contains no free parameter, no assumed angle and no light-curve fit. From Table~\ref{tab:in}, $E_{\rm p,i}^{(A)}/E_{\rm p,i}^{(B)}=36.4\pm9.8$ and $E_{\rm iso}^{(A)}/E_{\rm iso}^{(B)}=(7.45\pm1.41)\times10^{5}$, so
\begin{equation}\label{eq:result}
n=3.76\pm0.29 ,\qquad \epsilon\propto\Gamma^{3.76\pm0.29} .
\end{equation}
Integrating Eq.~(\ref{eq:master}) numerically for synthetic jets of known $k$ recovers $n=k$ to $0.005$ at the measured value, sixty times below the uncertainty; \cite[see][]{SM}. That uncertainty propagates the four measured quantities of Table~\ref{tab:in} and nothing else. It does not cover the premise, which two bursts cannot test, so what follows is conditional on it, as structures obtained by fitting an afterglow are conditional on a circumburst density and a set of microphysical parameters.

\begin{figure}[t]
\includegraphics[width=\columnwidth]{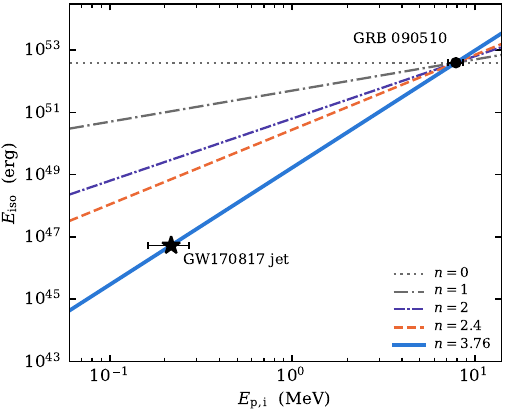}
\caption{\label{fig:meas}The two bursts in the $E_{\rm p,i}$--$E_{\rm iso}$ plane. The solid
line is the measured $n=3.76$. The others are the structures it excludes.}
\end{figure}

\textit{What it excludes.}---Four structures in common use follow definite values of $n$ (see Fig.~\ref{fig:meas}). An outflow with no energy structure gives $n=0$ and fails at $13\sigma$. One structured only in its baryon load gives $n=1$ and fails at $9.6\sigma$. The minimal structured jet, in which the angular dependence of the radiated energy comes entirely from the Lorentz boost of a uniform comoving flow \cite{2015MNRAS.450.3549S}, gives $n=2$ and fails at $6.1\sigma$. A uniform cone seen from just outside its edge gives $n=2.4$ when Eq.~(\ref{eq:master}) is integrated for a top-hat, or $2.0$--$2.3$ in the analytic and fitted treatments of \cite{2018PTEP.2018d3E02I,2017ApJ...850L..24G}. The largest of these fails at $4.7\sigma$. Carrying the full range of peak energies reported for GW170817, the first three remain excluded above $4\sigma$, the minimal structured jet, which is the closest of the four to the structures now in use, at $4.0\sigma$ or better, while the uniform cone falls to $2.8\sigma$ \cite{SM}.

We note that the classical point-source reading is not excluded by the energetics. With $E_{\rm iso}\propto\delta^4$ and $E_{\rm p}\propto\delta$ it predicts $n=4$, which Eq.~(\ref{eq:result}) matches to $0.8\sigma$. The geometry excludes it. Material whose velocity makes an angle $\psi$ with the line of sight has $\delta\le1/\sin\psi$ whatever its Lorentz factor, so at $\psi=\theta_v=20^\circ$ no emitter reaches $\delta=2.9$, while matching GRB 090510 on axis requires $\delta=82$ \cite{SM}.

\textit{The structure.}---We fix the energy structure relative to the Lorentz factor structure using Eq.~(\ref{eq:result}). Converting it into a run with polar angle is the one step that uses the anchors of Table~\ref{tab:in}. Equation (\ref{eq:Ep}) gives $\Gamma(\theta_v)=\Gamma_c/36.4=33$, and with a profile of the form used by \citet{2019Sci...363..968G}, $\Gamma(\theta)=1+(\Gamma_c-1)[1+(\theta/\theta_c)^2]^{-a/2}$, the index $a$ follows from $\theta_c$ and $\theta_v$ alone:
\begin{equation}\label{eq:answer}
\theta_c=2^\circ\textrm{--}5^\circ,\quad a=1.6\textrm{--}2.6,\quad \Gamma(\theta_v)=33 ,
\end{equation}
bracketing the canonical $\Gamma-1\propto\theta^{-2}$ without that value being imposed. The index is quoted at the fiducial $\theta_v=20^\circ$; carrying the full interferometric range $19^\circ$--$25^\circ$ broadens it to $a=1.4$--$2.7$ \cite{SM}. The radiated structure outside the core is $\epsilon\propto\theta^{-7.4}$ for the fiducial $a=1.90$, using the band corrected exponent $\epsilon\propto\Gamma^{3.88}$ \cite{SM}.

\begin{figure}[t]
\includegraphics[width=\columnwidth]{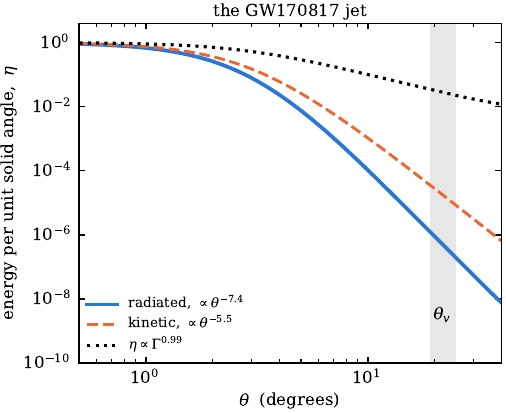}
\caption{\label{fig:struct}The radiated energy per unit solid angle measured here, the
kinetic energy per unit solid angle from the afterglow \cite{2019Sci...363..968G}, and their
ratio, the prompt radiative efficiency. The band marks the interferometric viewing angle.}
\end{figure}

\textit{The efficiency.}---The afterglow of the same jet measures a different quantity, the kinetic energy per unit solid angle that remained to drive the external shock. Its ratio to the radiated energy is the prompt radiative efficiency $\eta$, which \citet{2015MNRAS.450.3549S} identified as unknown and possibly angle dependent. Taking their determination with ours (Fig.~\ref{fig:struct}),
\begin{equation}\label{eq:eta}
\eta(\theta)=\frac{\epsilon(\theta)}{{\rm d}E_{\rm kin}/{\rm d}\Omega}
\propto\Gamma^{0.99\pm0.29} ,
\end{equation}
so the prompt emission converts energy to gamma-rays in direct proportion to the local Lorentz factor. Between the core and the line of sight the efficiency falls by a factor $35$. This is not a consistency check between two measurements of one quantity but a determination of a third from two independent ones. It is also the one result that inherits quantities from elsewhere, and it inherits them twice: through the angular anchor, which enters via $a$, and through the afterglow slope itself, which \citet{2019Sci...363..968G} determine as $5.5$ with a one-sigma range $4.1$--$6.8$. Propagating both, the exponent runs from $-1.0$ to $2.4$. The rise is what the fiducial values give, and it is not established by them \cite{SM}.

\begin{figure}[t]
\includegraphics[width=\columnwidth]{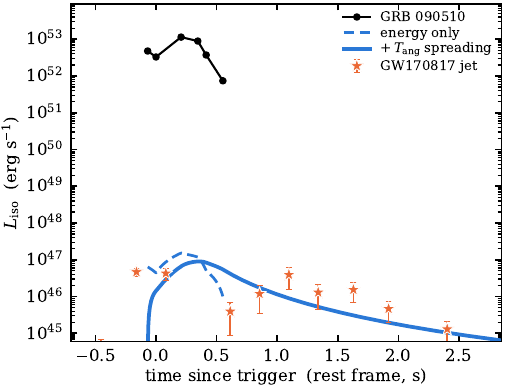}
\caption{\label{fig:lc}The light curve of GRB 090510 carried to $\theta_v$. Dashed: scaled by
the measured energy ratio alone. Solid: the same curve convolved with the high-latitude
response of width $T_{\rm ang}=0.41$~s, which conserves the energy. Stars: GW170817 as
observed. Data from \cite{2010ApJ...716.1178A,2017ApJ...848L..14G}.}
\end{figure}

\textit{Timing.}---For a shell radiating at $R_\gamma$ with Lorentz factor $\Gamma$ along the line of sight the arrival times from the visible patch are spread by $T_{\rm ang}=R_\gamma/2\Gamma^2c$. The main gamma-ray pulse of GW170817 is longer than that of GRB 090510 by a factor $3.6$ in the rest frame. We attribute the widening to the angular spreading alone, which gives the two engines the same duration and is part of treating the two as systems of the same kind. This leads to $T_{\rm ang}=0.41$~s and
\begin{equation}\label{eq:R}
R_\gamma=2\Gamma^2(\theta_v)c\,T_{\rm ang}=2.7\times10^{13}~{\rm cm} .
\end{equation}
The same radius contributes $0.41$~s to the lag of the gamma-rays behind the merger, leaving $1.31$~s in the source frame for jet launch and breakout, which matches the published decomposition. We find the spreading to be required by the light curve itself. Scaling the rest-frame light curve of GRB 090510 to $\theta_v$ by the measured energy ratio alone leaves a pulse too high by a factor $3.3$ and too narrow; convolving it with the high-latitude response of width $T_{\rm ang}$, which conserves energy, lowers the peak by a factor of $1.7$ and reproduces the observed decay, $\chi^2/N=1.0$ beyond $0.7$~s (Fig.~\ref{fig:lc}). Radiating material genuinely $20^\circ$ from the line of sight would carry a path term of $54$~s and is therefore excluded. At this radius the shell is transparent, $\tau=2.0\times10^{-4}/\eta(\theta_v)$. Since $R_\gamma=2\Gamma^2cT_{\rm ang}$, the optical depth scales as $\Gamma^{-5}$, and we find that transparency alone requires $\Gamma_c\gtrsim365$. This bound does not use the compactness limit, but neither is it independent of what precedes it, since the radius it rests on follows from the pulse widening. It should be viewed as a condition the picture must meet rather than a constraint imposed from outside \cite{SM}.

\textit{Conclusion.}---The radiated angular structure of a short-burst jet is accessible from two prompt spectra, with no afterglow, no circumburst density, and no microphysical parameter. Combined with the afterglow it gives the prompt radiative efficiency, which neither measurement determines by itself. Two bursts define a slope and no more, a line through two points leaves no residual. A third burst with an independently measured Lorentz factor would convert the slope of Eq.~(\ref{eq:result}) into a shape, and it would be the first test capable of falsifying the premise.
The result supports the gravitational-wave emission recorded by
LIGO--Virgo \cite{2017PhRvL.119p1101A} and the jet emission of GW170817 measured by
\citet{2018Natur.561..355M,2022Natur.610..273M}: from the prompt emission alone we
obtain a line-of-sight Lorentz factor $\Gamma(\theta_v)=33$ and a core exceeding the
$\Gamma_i>40$ their optical astrometry requires \cite{2022Natur.610..273M}, both well above
the $\Gamma\ge4.1$ that the superluminal motion itself demands
\cite{2018Natur.561..355M}.

Throughout we have proceeded from the premise that GRB 090510 and GW170817 are systems of the same kind, seen from two different directions.

\begin{acknowledgments}
We thank our colleagues at ICRANet for discussions. In preparing this manuscript the authors used Claude (Anthropic) for language editing and for checking the  internal consistency of the derivations. The analysis, the results and the conclusions are the authors' own, and the authors take full responsibility for the content of this work. The analysis was performed by NumPy \cite{2020Natur.585..357H} and Matplotlib \cite{2007CSE.....9...90H}.
\end{acknowledgments}

\bibliographystyle{apsrev4-2}
%

\clearpage
\onecolumngrid
\begin{center}
\textbf{\large Supplemental Material}\\[4pt]
\textbf{The angular structure of the GW170817 jet from prompt emission alone}
\end{center}
\vspace{10pt}
\setcounter{section}{0}
\setcounter{equation}{0}
\setcounter{figure}{0}
\setcounter{table}{0}
\renewcommand{\thesection}{S\arabic{section}}
\renewcommand{\theequation}{S\arabic{equation}}
\renewcommand{\thefigure}{S\arabic{figure}}
\renewcommand{\thetable}{S\arabic{table}}

\section{The error budget of the filled-cone limit}

The kernel of Eq.~(\ref{eq:master}) integrates to $4\pi$ exactly when $\Gamma$ does not depend on $\theta$, the integral running over the full solid angle. A quarter of it originates from outside the nominal beaming cone: within $1/\Gamma$ the integral reaches $75\%$ of $4\pi$, within $2/\Gamma$ it reaches $96\%$, and within $3/\Gamma$, $99\%$. The cancellation is therefore a property of the whole sphere and not of the cone alone.
 
Once $\Gamma$ depends on $\theta$ the kernel is no longer $4\pi$. For the profile adopted from \cite{2019Sci...363..968G}, in which $\Gamma-1$ falls as $\theta^{-a}$ outside a core of half-angle $\theta_c$, it exceeds that value by $1.5\%$ on axis and $4.8\%$ at $\theta_v=20^\circ$, rising to $66\%$ if the index is steepened to $a=2.5$. Nor is $\epsilon$ constant across the visible patch, which at $\theta_v=20^\circ$ spans $1/\Gamma(\theta_v)=1.7^\circ$, over which $\Gamma$ changes by a factor $1.4$ and $\epsilon$, falling as $\Gamma^{3.76}$, by a factor $3.3$. The two departures together leave Eq.~(\ref{eq:Eiso}) in error by $0.04\%$ on axis and $1.9\%$ at $\theta_v$. The size is controlled by the ratio of the patch width to the angular scale on which the profile varies, $a/[\Gamma(\theta_v)\theta_v]$, which is $0.17$ for the solution obtained. The departure is second order in that ratio. Over $a=1.0$--$2.5$ the ratio runs from $0.016$ to $0.64$, $\epsilon$ varying across the patch by factors from $1.1$ to $81$, and the departure stays between $0.64$ and $0.73$ times its square, giving $0.02\%$, $0.3\%$, $1.9\%$, $7.5\%$ and $26\%$. The limit is therefore one of smoothness and not of uniformity.

An error of $1.9\%$ in $E_{\rm iso}$ propagates to $0.019/\ln 36.4=0.005$ in $n$. This agrees with the direct test. Constructing synthetic jets with $\epsilon\propto\Gamma^k$ and integrating Eq.~(\ref{eq:master}) over the whole outflow, with no expansion anywhere, recovers $n=-0.009$, $0.998$, $2.000$, $2.998$, $3.755$, $4.985$ and $5.971$ for $k=0$, $1$, $2$, $3$, $3.76$, $5$ and $6$. The largest error is $0.03$ and that at the measured value is $0.005$, fifty times below the statistical uncertainty on $n$. The test uses a known input and never uses Eq.~(\ref{eq:Eiso}), so it is not circular. Repeating it with steeper profiles, the recovered exponent is accurate to $0.05$ for $a\le2.5$ and degrades to $0.3$ by $a=3.5$, where $\Gamma(\theta_v)$ has fallen to $2.5$ and the beaming patch is no longer narrow. The solution obtained has $a=1.4$--$2.7$; at its steep end the recovered exponent is in error by about $0.09$, still well below the statistical uncertainty on $n$.

\section{Conditions under which Eq.~(\ref{eq:master}) holds, and the transparency bound}

Equation~(\ref{eq:master}) applies for an emitting region that is far from the observer, geometrically thin, transparent, and whose radius does not change appreciably while it radiates, with an emissivity isotropic in the comoving frame. With $R_\gamma$ determined the first four can be checked. The source is distant: $d_L=42$~Mpc at $z=0.0099$ is $5\times10^{12}$ emission radii. The shell is thin: after subtracting the angular spreading from the rest-frame main pulse, the remaining engine time is, $T_{\rm eng}=0.570-0.41=0.16$~s, leading to a radial width $\Delta=cT_{\rm eng}=4.8\times10^{9}$~cm, so $\Delta/R_\gamma=1.8\times10^{-4}$, and the radius over which a given layer radiates changes by the same fraction.

The shell is optically thin to Thomson scattering. Its isotropic-equivalent mass is $M=E_{\rm iso}/[\eta\,\Gamma(\theta_v)c^2]$, and we obtain for the optical depth through the radial column
\begin{equation}
\tau_{\rm T}=\frac{\sigma_{\rm T}M}{4\pi R_\gamma^2m_p}
=\frac{\sigma_{\rm T}E_{\rm iso}}
{4\pi\,\eta\,\Gamma(\theta_v)\,m_pc^2R_\gamma^2}
=\frac{7.8\times10^{-5}}{\eta(\theta_v)} .
\end{equation}
No engine duration enters: spreading changes the shell width and its density together and leaves the column unaltered. The suppression of the scattering rate by $1-\beta$ for a photon overtaking the flow is cancelled by the $1/(1-\beta)$ longer path it must traverse to leave the shell, so the lab-frame and comoving estimates coincide. At $\eta(\theta_v)=0.03$ the  optical depth is $2.6\times10^{-3}$, so the shell is Thomson thin at the radius where the gamma-rays are released. Because $R_\gamma$ is itself set by $\Gamma(\theta_v)$ through Eq.~(\ref{eq:R}), carrying that dependence through leaves $\tau\propto\Gamma^{-5}$, and transparency alone requires $\Gamma(\theta_v)\gtrsim10$, that is $\Gamma_c\gtrsim365$. The bound is insensitive to the efficiency it contains, since the threshold scales as $\eta^{-1/5}$: over the two decades $\eta=0.003$--$0.3$ varies only between $\Gamma_c\gtrsim580$ and $\gtrsim230$.

The fifth condition, isotropy of the comoving emissivity, cannot be tested. It enters weakly because Eq.~(\ref{eq:def}) is a ratio between two bursts already assumed to be of the same kind: an anisotropy common to the pair cancels, and only a difference between them would shift $n$. The same applies for the geometric factors $R^2\Delta$ in the decomposition $\epsilon=\eta R^2\Delta u$ with $u=\Gamma^2u'$, which are absorbed into $\epsilon$.

\section{The band correction}

Equation~(\ref{eq:def}) relates bolometric quantities, whereas the two energies are band limited and over different bands, so a cosmological $k$-correction is  required \cite{2001AJ....121.2879B}. The isotropic energy within an observed band is the integral of the structured-jet spectrum over that band. Evaluating it for the structure derived, with the same comoving spectrum for both bursts, the rest-frame window $1$--$10^4$~keV contains $36\%$
of the bolometric energy of GRB 090510, whose peak lies near its upper edge, while
$10.1$--$1010$~keV contains $48\%$ for GW170817. The displacement is
$n_{\rm bol}-n=+0.12$, so $\epsilon\propto\Gamma^{3.88}$ and $\epsilon\propto\theta^{-7.4}$. The result is stable: input slopes $3.4$, $3.76$ and $4.1$ all return $+0.119$. It is smaller than a Band-function estimate applied to each burst separately would give, because the spectrum seen by an off axis observer is a superposition over the beaming patch and is broader than any single Band function.

\section{The light-curve transformation}

We construct Fig.~\ref{fig:lc} as follows. The rest-frame light curve of GRB 090510 is taken as the upper envelope of the Fermi-GBM record and normalised so that its integral is the isotropic energy of Table~\ref{tab:in}, $3.95\times10^{52}$~erg; that of GW170817 is normalised in the same way to $5.3\times10^{46}$~erg. Both are in the source frame, so no time dilation enters.

The dashed curve is the first divided by the measured energy ratio $E_{\rm iso}^{(A)}/E_{\rm iso}^{(B)}=7.45\times10^5$ and no other correction. It peaks at $1.53\times10^{47}$~erg~s$^{-1}$, a factor $3.3$ above the observed peak of $4.66\times10^{46}$~erg~s$^{-1}$, and is too narrow.

The solid curve adds the high-latitude response. We derive its kernel from Eq.~(\ref{eq:master}) rather than assuming it. Material at an angle $\psi$ from the line of sight is seen delayed by $t=R_\gamma\psi^2/2c=T_{\rm ang}(\Gamma\psi)^2$ and carries the Doppler factor $\delta=2\Gamma/(1+t/T_{\rm ang})$. The solid angle reaching the observer per unit time, ${\rm d}\Omega/{\rm d}t=2\pi c/R_\gamma$, is constant, so the integrand of Eq.~(\ref{eq:master}) gives ${\rm d}E/{\rm d}t\propto\delta^3$ and the arrival-time kernel is
\begin{equation}
k(t)=\frac{2}{T_{\rm ang}}\left(1+\frac{t}{T_{\rm ang}}\right)^{-3},
\qquad t\ge0,
\end{equation}
with $T_{\rm ang}=0.41$~s, the value fixed by the pulse widening. We verify that the index is not an artifact of taking the patch uniform. Integrating Eq.~(\ref{eq:master}) over the structure itself, with $\Gamma$ and $\epsilon$ varying across the patch as the profile requires, we obtain $3.01$ at the fiducial geometry and $3.01$--$3.06$ over the whole range of viewing angles and core widths admitted here. The kernel integrates to unity, so the convolution conserves the energy and only redistributes it in time. The convolved curve peaks at $9.1\times10^{46}$~erg~s$^{-1}$, a factor $1.7$ below the unspread curve and within a factor $1.9$ of the observed peak. The decay is matched more closely than the peak, as it should be, since it is there that the high-latitude response dominates: over the points beyond $0.7$~s the convolved curve gives $\chi^2/N=1.0$. No parameters are adjusted; the energy ratio is measured and $T_{\rm ang}$ follows from the widening of the main pulse.

\section{Systematics}

\textit{Spectral component.}---The peak energy adopted for GW170817 belongs to the Comptonized component, whereas the isotropic energy is the sum of that component and the blackbody tail. Taking both from the Comptonized component alone raises $n$ by $0.08$, a quarter of the statistical uncertainty.

\textit{Peak energy.}---Reported values of $E_{\rm peak}$ for GW170817 span $82$ to $215$~keV. Table~\ref{tab:syst} propagates the range, holding $E_{\rm iso}$ at the whole-burst value so that the peak energy alone varies. The exponent and the line-of-sight Lorentz factor change substantially. The exclusions change much less, because the uncertainty on the exponent falls with it, as the table shows. Over the whole range the four structures of the Letter are excluded at $10.3$--$16.2\sigma$, $7.2$--$10.7\sigma$, $4.0$--$6.1\sigma$ and $2.8$--$4.7\sigma$. Three of the four are therefore excluded above $4\sigma$ at every entry. The fourth is not: the exponent exceeds the uniform-cone value everywhere, but by only $2.8\sigma$ at the entry carrying the largest reported error, $128\pm49$~keV. It is on that structure alone that the peak energy adopted for GW170817 bears.

\begin{table}[h]
\caption{\label{tab:syst}Effect of the adopted peak energy of GW170817. The index $a$ is evaluated at $\theta_c=3^\circ$, $\theta_v=20^\circ$, and $q=n+0.12 5.5/a$ is the implied efficiency index. The uncertainty $\sigma_n$ propagates the four observables of Table~\ref{tab:in} at each entry.}
\begin{ruledtabular}
\begin{tabular}{rccccc}
$E_{\rm peak}$ (keV, obs.) & $n$ & $\sigma_n$ & $\Gamma(\theta_v)$ & $a$ & $q$ \\
\hline
$82\pm21$  & $2.97$ & $0.18$ & $12.6$ & $2.43$ & $0.82$ \\
$128\pm49$ & $3.29$ & $0.32$ & $19.7$ & $2.18$ & $0.89$ \\
$185\pm62$ & $3.61$ & $0.34$ & $28.4$ & $1.98$ & $0.95$ \\
$215\pm54$ & $3.76$ & $0.29$ & $33.0$ & $1.90$ & $0.99$ \\
\end{tabular}
\end{ruledtabular}
\end{table}

\textit{Comoving spectrum.}---Equation~(\ref{eq:Ep}) is obtained for one comoving shape applied to both bursts. With the same shape at both angles the coefficient cancels in the ratio to $0.03\%$ whatever the indices, so the relation does not depend on the shape adopted, only on the two bursts sharing it. A mismatch does not cancel: varying the assumed comoving indices independently over the range short-burst spectra span displaces $n$ by up to $0.16$, about half its uncertainty. The observed fits are not even of one functional form, Comptonized for GW170817 and Band for GRB 090510, so the assumption is not one the data confirm. This is one of the things assumed when the two are treated as systems of the same kind.

\textit{Viewing angle.}---The published determinations differ, $\theta_v\simeq20^\circ$ and $19^\circ$--$25^\circ$ from astrometry and $15^{+1.5}_{-1.0}$ degrees from a joint fit of the afterglow and the centroid motion. Equations~(\ref{eq:def}) and (\ref{eq:result}) contain no angle, so the exponent and the exclusions are untouched. The angle enters only the conversion to a run with polar angle, through the ratio $\theta_v/\theta_c$. Both $\Gamma_c$ and $\Gamma(\theta_v)$ are obtained without reference to any angle, so solving the profile for $\theta$ at $\theta=\theta_v$ returns not the viewing angle but
\begin{equation}\label{eq:ratio}
\frac{\theta_v}{\theta_c}=\left[\left(\frac{\Gamma_c-1}{\Gamma(\theta_v)-1}\right)^{2/a}
-1\right]^{1/2} .
\end{equation}
There is no angular scale in two prompt spectra, and the geometry is fixed only  up to that number. Adopting the canonical $a=2$ in place of the astrometry gives $\theta_v/\theta_c\simeq6$, so that the interferometric angle predicts a core width $\theta_c\simeq3^\circ$--$4^\circ$, inside the limit $\theta_c<5^\circ$.

\textit{Core Lorentz factor.}---Little of the structure depends on where in the allowed range $\Gamma_c$ falls. Reducing it from $1200$ to the transparency floor of $365$ carries $a$ from $1.90$ to $1.93$ and the radiated run from $\theta^{-7.4}$ to $\theta^{-7.5}$. Raising it has even less effect, which is the relevant direction since the compactness limit is a lower bound: Eq.~(\ref{eq:Ep}) ties $\Gamma(\theta_v)$ to $\Gamma_c$, so the ratio $(\Gamma_c-1)/(\Gamma(\theta_v)-1)$ that fixes $a$ is set by the measured peak-energy ratio and not by the anchor, and a decade in $\Gamma_c$, to $10^4$, moves $a$ only from $1.90$ to $1.89$. What the anchor fixes is $\Gamma(\theta_v)$ and, through it, the emission radius. The afterglow fit of \citet{2019Sci...363..968G}, whose energy slope we adopt, returns for this jet $\log\Gamma_c=2.4$ with a one-sigma range $2.0$--$2.9$, and a Lorentz-factor index $s_2=3.5$ with a range $1.8$--$5.6$. The core adopted here is faster than their fit returns and the index shallower, meeting their range only at its lower end. The two determinations are independent, theirs resting on the afterglow and ours on the two prompt spectra, and we report the disagreement rather than adjust to it. It does not propagate into $a$, which $\Gamma_c=250$ moves only to $1.96$, but their central value lies below the transparency floor obtained above, so the picture requires a core faster than the afterglow fit returns.

\textit{Efficiency.}---The efficiency index $q=n+0.12-s_1/a$ depends on quantities taken from elsewhere more deeply than the exponent does, because the afterglow slope $s_1$ is a run with polar angle and must be carried through $a$. At the fiducial $\theta_v=20^\circ$, with $s_1=5.5$, the index is $a=1.6$--$2.6$ and $q=0.44$--$1.77$. Carrying the full interferometric range gives $a=1.4$--$2.7$ and $q=-0.05$--$1.84$. The slope itself is not exact either: \citet{2019Sci...363..968G} determine $s_1=5.5$ with a one-sigma range $4.1$--$6.8$, which at $a=1.90$ alone gives $q=0.30$--$1.72$, and over both ranges together $q=-0.98$--$2.36$. At the fiducial values $q=0.99\pm0.29$ excludes a constant efficiency at $3.4\sigma$ and the efficiency rises with the Lorentz factor; over the ranges above the sign of $q$ is not determined. What the two determinations fix well is that the radiated and the kinetic structures are not the same function of angle.

\section{The point-source reading}

With $E_{\rm iso}\propto\delta^4$ and $E_{\rm p}\propto\delta$ the classical reading predicts $n=4$, which the measurement matches to $0.8\sigma$, and the Doppler ratio it requires, $(E_{\rm iso}^{(A)}/E_{\rm iso}^{(B)})^{1/4}=29.4$, agrees with the observed peak-energy ratio to $0.7\sigma$. The energetics do not separate it from the reading adopted.

The geometry, however, does. Whatever its Lorentz factor, material whose velocity makes an angle $\psi$ with the line of sight has $\delta\le1/\sin\psi$, the maximum reached at $\Gamma=1/\sin\psi$; at $\psi=\theta_v=20^\circ$ no emitter exceeds $\delta=2.9$. Matching GRB 090510 on axis, where $\delta=2\Gamma_c$, requires $\delta=82$ for GW170817, a factor $28$ beyond what the geometry allows. The reading survives only for $\Gamma_c\le43$, against the $\gtrsim1200$ of the compactness limit and the $\gtrsim365$ obtained above. The bound binds only because a point source is offset from the line of sight; for the structured outflow the radiating material lies on it and $\delta=2\Gamma(\theta_v)$ is unconstrained by it.

\section{The gravitational-wave comparison}

The total energy a binary radiates in gravitational waves does not depend on the direction from which it is seen. It is a property of the system, and the waveform analysis solves for $\iota$ in extracting it. Two binaries of the same kind therefore release the same gravitational-wave energy at any inclination, and a comparison between them is not affected by the angle that separates their gamma-rays by six decades.

Applying to the binary of GRB 090510 the same treatment of its prompt emission, \citet{2026JHEAp..5000464R} infer from the electromagnetic observables alone a release of $\Delta E_{\rm insp}<0.018\,Mc^2=7.6\times10^{52}$~erg during the  inspiral and $\Delta E_{\rm mgr}\approx0.007\,Mc^2=2.96\times10^{52}$~erg at merger, radiated near $2$~kHz. For GW170817 the corresponding measurement is a lower limit on the energy emitted below $600$~Hz, $E_{\rm rad}>0.025\,M_\odot c^2=4.5\times10^{52}$~erg \citep{2017PhRvL.119p1101A}, obtained for a binary of total mass $2.74^{+0.04}_{-0.01}\,M_\odot$ against the $2.38\,M_\odot$ inferred for GRB 090510. The two limits are compatible. They do not cover the same range: the bound for GW170817 stops at $600$~Hz, whereas the binding-energy difference above runs to the merger radius, near $2$~kHz, so it is a lower limit on part of what the other bounds from above. Nothing was recovered at the merger frequency itself, the detectors losing sensitivity above a kilohertz. The same analysis reports the inclination as a separate parameter, $\Theta\le28^\circ$ using the position of NGC 4993, so neither energy carries angular information. None of the three is a two-sided measurement, so no difference between them can be formed; what the comparison establishes is consistency, not a match.

The gravitational-wave strain alone cannot determine the viewing angle independently of the luminosity distance, and the interferometric constraint is therefore used as the angular anchor. At leading quadrupole order a circular binary radiates $h_+\propto(1+\cos^2\iota)/2$ and $h_\times\propto\cos\iota$ \citep{1963PhRv..131..435P,2009LRR....12....2S}, so the energy per unit solid angle falls by a factor of $8$ from pole to equator and by no more. At $\iota=20^\circ$ the amplitude is $0.941$ of its face-on value and the energy per unit solid angle $0.885$, reductions of $5.9\%$ and $11.5\%$. The amplitude is degenerate with distance \citep{1993PhRvD..47.2198F,2019ApJ...877...82U}, which is why the waveform alone gave $40^{+8}_{-14}$~Mpc. The angular anchor is therefore taken from the interferometry and not from the strain. Because detectable volume scales as the cube of the amplitude, gravitational-wave selection favours face-on systems only mildly, raising the share of detections within $20^\circ$ of the axis from $6.0\%$ to $19\%$ \citep{2011CQGra..28l5023S}, so most mergers of this kind should be recorded with no detectable prompt counterpart.

\section{Relation to earlier off-axis treatments}

That an off-axis viewer sees a burst with $E_{\rm p}\propto\delta$ and a strongly suppressed isotropic energy was developed by \citet{2001ApJ...554L.163I} and applied to X-ray flashes by \citet{2002ApJ...571L..31Y,2003ApJ...593..941Y,2003ApJ...594L..79Y}, with the afterglow treatment of \citet{2002ApJ...570L..61G} and the structured-jet framework of \citet{2002MNRAS.332..945R,2002ApJ...571..876Z}. What separates the present treatment from all of these is that the beaming cone is filled, so no de-beaming factor is applied.

\citet{2017ApJ...850L..24G} inferred an on-axis peak energy of $3$--$8$~MeV for GW170817 and remarked that such a value is unusually high for a short burst of ordinary isotropic energy. That inference is made at $n=2$: their relations are written for a jet with sharp edges viewed at $1<\theta_v/\theta_j<2$, where $E_{\rm iso}$ falls as the square of the factor by which $E_{\rm p}$ falls. Reproducing the observed energy ratio requires a peak-energy ratio of $863$ at $n=2$ but only $36.5$ at $n=3.76$, a factor $24$ smaller. Repeating their calculation with the measured exponent and their own assumed on-axis energies, $10^{49}$--$10^{51}$~erg, returns $0.87$--$2.9$~MeV, a factor $1.4$--$5.9$ above the $0.5$--$0.6$~MeV they quote as typical, in place of the factor $4.9$--$59$ that $n=2$ requires. Finding the sharp-edged case untenable, those authors preferred one in which the jet has no sharp edges and the prompt emission comes from outflow along the line of sight at $\theta_v\gtrsim2\theta_j$, which is the configuration adopted here.

They also used the delay to bound the emission radius, $R_\gamma<2c\Delta t/\Delta\theta^2$, obtaining $R_\gamma\lesssim1.7\times10^{12}(\Delta\theta/0.25)^{-2}$~cm. The bound assumes the radiating material is offset from the line of sight, and does not apply when it lies on it.

\citet{2018PTEP.2018d3E02I} reached the same conclusion for the energetics, obtaining $2.3$ by fitting the envelope of a numerically integrated top-hat, and question the compactness limit for GRB 090510 as resting on a one-zone treatment. The transparency bound above bears on that: a core Lorentz factor reduced to $\sim300$ would put $\Gamma(\theta_v)$ near $8$ and, since $R_\gamma$ falls as $\Gamma^2$, would leave the emitting shell optically thick, which the observed non-thermal spectrum excludes.

Other origins have been advanced for the prompt emission of GW170817, principally cocoon breakout \cite{2018MNRAS.479..588G,2018MNRAS.475.2971B,2018ApJ...867...18N} and structured-jet models fitted to the full data set \cite{2018MNRAS.481.1597G,2018MNRAS.478..733L}; a photospheric fit gives a line of-sight Lorentz factor of $\sim20$ \cite{2018ApJ...860...72M}, and \citet{2019A&A...628A..18S} inverted the problem, using the structure inferred from the afterglow to compute how the burst would appear on axis. We do not adjudicate between them. The measurement reported here uses only two prompt spectra and the interferometric geometry, and is therefore testable against the wider body of data \cite{2017ApJ...848L..15S,2017Natur.551...71T,2017Sci...358.1579H,2021ApJ...922..154M} without having been fitted to it. We find the core width obtained here to agree with independent determinations of short-GRB jet structure \cite{2019MNRAS.482.5430B}: \citet{2023A&A...680A..45S} obtain $\theta_c=2.1^{+2.4}_{-1.4}$ degrees from $367$ short bursts, and simulations of GRB 090510 itself \cite{2025A&A...702A..13S} reproduce its energetics and opening angle. The nature of the remnant of GRB 090510 is itself unsettled: it is a black hole in the treatment adopted by \citet{2026JHEAp..5000464R}, whereas the plateau in its X-ray afterglow has been fitted with a stable magnetar \cite{2013MNRAS.430.1061R}. Nothing here turns on the difference, since the exponent is measured from two prompt spectra and the afterglow of GRB 090510 enters only as the source of its redshift. Finally, where the rest-frame energy of a spectral feature is known independently, the scaling $E_{\rm p}\propto\Gamma$ of Eq.~(\ref{eq:Ep}) is measurable rather than assumed: the line seen in GRB 221009A drifting from $\sim37$ to $\sim6$~MeV has been identified with the $158$~keV decay branch of $^{56}$Ni entrained in the jet \cite{2026CmPhy...9..172M}, tracking $\delta$ from $\sim230$ to $\sim40$ within a single burst.

\end{document}